\documentclass[11pt]{article}
\usepackage[dvips]{graphicx}
\usepackage{latexsym}
\usepackage{amssymb}

\newcommand{\CP}{\mathbb{CP}}
\newcommand{\RP}{\mathbb{RP}}
\newcommand{\R}{\mathbb{R}}

\renewcommand{\d}{\mathrm{d}}

\def\be{\begin{equation}}
\def\ee{\end{equation}}

\def\Sm{\Sigma}

\def\Om{\Omega}

\def\om{\omega}
\def\ov{\overline}

\def\p{\partial}

\def\ov{\overline}

\def\l{\lambda}

\def\O{{\cal O}}

\def\u{\omega}

\newtheorem{prop}{Proposition}  

\begin{document}
\title{Hyper-complex four-manifolds from the Tzitz\'eica  equation}
\author{Maciej  Dunajski\thanks{email: dunajski@maths.ox.ac.uk}
\\ The Mathematical Institute,
24-29 St Giles, Oxford OX1 3LB, UK}  
\date{} 
\maketitle
\abstract{It is shown how solutions to the Tzitz\'eica  equation
can be used to construct a family of (pseudo) hyper-complex metrics
in four dimensions.} 
\section{Introduction}
A striking {\it universal} feature of integrable systems is that the same integrable equations  often arise from many unrelated sources. 
The Tzitz\'eica equation \cite{Tz08}
\be
\label{tzitzeica}
\u_{xy}=e^\u-e^{-2\u}
\ee
is a good example. It first arose in a study of surfaces in $\R^3$ 
for which the ratio of the negative Gaussian curvature to the fourth power of
a distance from a tangent plane to some fixed point is a constant.
Tzitz\'eica has shown that if $x$ and $y$ are coordinates on such a
surface in which the second fundamental form is off-diagonal, then
there exists a real function $\u(x,y)$ such that the Peterson-Codazzi
equations reduce to (\ref{tzitzeica}). Moreover, he has demonstrated 
\cite{Tz10}
that (\ref{tzitzeica}) is a consistency condition for an otherwise 
overdetermined system of PDEs\footnote{Strictly speaking the linear system 
given by Tzitz\'eica consisted of three second order PDEs for one
function. These three equations can be recovered from (\ref{T10})
if one eliminates $\psi_1$ and $\psi_2$ by cross-differentiating.}
for $\psi_i(x, y), \;i=1, 2, 3.$
\begin{eqnarray}
\label{T10}
\partial_x\left(\begin{array}{c}
{{\mbox{\boldmath $\psi$}}}_1\\
{{\mbox{\boldmath $\psi$}}}_2\\
{{\mbox{\boldmath $\psi$}}}_3
\end{array}\right)&=&
\left(\begin{array}{ccc}
-\u_x& 0&\l\\
1 & \u_x &0\\
0&1&0
\end{array}\right)
\left(\begin{array}{c}
{{\mbox{\boldmath $\psi$}}}_1\\
{{\mbox{\boldmath $\psi$}}}_2\\
{{\mbox{\boldmath $\psi$}}}_3
\end{array}\right),\\
\partial_y\left(\begin{array}{c}
{{\mbox{\boldmath $\psi$}}}_1\\
{{\mbox{\boldmath $\psi$}}}_2\\
{{\mbox{\boldmath $\psi$}}}_3
\end{array}\right)&=&
\left(\begin{array}{ccc}
0&e^{-2\u}&0\\
0 &0&e^\u\\
\l^{-1}e^\u&0&0
\end{array}\right)\left(
\begin{array}{c}
{{\mbox{\boldmath $\psi$}}}_1\\
{{\mbox{\boldmath $\psi$}}}_2\\
{{\mbox{\boldmath $\psi$}}}_3
\end{array}\right)\nonumber.
\end{eqnarray}
The above linear system  is in modern terminology known as a 
`Lax pair with a spectral parameter'. It 
underlines the complete integrability of the 
Tzitz\'eica  equation \cite{M81}. 

Equation (\ref{tzitzeica}) 
reappeared in the context of soliton
solutions \cite{DB77}, gas dynamics \cite{G84}
as well as  geometry of affine spheres \cite{J53}.
In this paper I shall reveal yet another occurrence of
(\ref{tzitzeica}), and show how its solutions  can be used to generate
explicit pseudo-hyper-complex structures in four dimensions.
This will be done by regarding (\ref{T10})
as a reduced Lax pair for $SL(3, \R)$ anti-self-dual Yang-Mills (ASDYM)
equations,  embedding $SL(3, \R)$ in Diff$(\RP^2)$, and reinterpreting
the Lax pair in terms of vector fields on ${\cal M}=\R^2\times \RP^2$.
Four independent vector fields in this Lax pair will provide
a null frame for a pseudo-hyper-complex conformal structure on $\cal M$.

In the next section the Lax formulation of the 
pseudo-hyper-complex condition in four dimensions will be given following
\cite{D99, GS99}. In \S\S\ref{S3} the connection with the ASDYM will 
be established, and the explicit embedding of $\mathbf{sl}(3, \R)$ in
$\mathbf{diff}(\RP^2)$ will be given. The resulting pseudo-hyper-complex
structure will be  constructed in \S\S\ref{S4}. All considerations in this
section will be local.
Finally \S\S\ref{S5}
contains the twistor interpretation of the construction.
\section{Pseudo hyper-complex structures}
\label{S2}
A smooth real $4n$-dimensional manifold ${\cal M}$ equipped with three real 
endomorphisms $I, S, T:T{\cal M}\rightarrow T{\cal M}$
of the tangent bundle 
satisfying the algebra of pseudo-quaternions 
\[
-I^2=S^2=T^2=1,\qquad IST=1,
\]
is called
pseudo-hyper-complex iff the almost complex structure
\be
\label{cpx}
{\cal J}_{\l}=aI+bS+cT
\ee
is integrable for any point of the hyperboloid\footnote{ We identify two sheets of this hyperboloid with two unit discs $D_-$ and $D_+$, and use $\l$ as a projective coordinate on a Riemann sphere $\CP^1=D_-+D_++S^1$. 
The coordinate $\l$ plays a role of a 
complex spectral parameter in the Lax pair (\ref{hclax}).}
$a^2-b^2-c^2=1$.
This integrability is equivalent to a vanishing of 
its Nijenhuis tensor
\[
N(X_1, X_2):=({\cal J_\l})^2[X_1, X_2]-
{\cal J}_\l [{\cal J}_\l X_1, X_2]-
{\cal J}_\l[X_1,{\cal J}_\l X_2]
+[{\cal J}_\l X_1,{\cal J}_\l X_2]
\]
for arbitrary vectors $X_1$ and $X_2$.
A convenient matrix representation of the canonical pseudo-hyper-complex 
structure 
on $\R^4$ is  given by
\[
I=\left (
\begin{array}{cccc}
0&0&1&0\\
0&0&0&1\\
-1&0&0&0\\
0&-1&0&0
\end{array}
\right ), \qquad S=\left (
\begin{array}{cccc}
0&0&1&0\\
0&0&0&1\\
1&0&0&0\\
0&1&0&0
\end{array}
\right ),\qquad
T=\left (
\begin{array}{cccc}
1&0&0&0\\
0&1&0&0\\
0&0&-1&0\\
0&0&0&-1
\end{array}
\right ).
\]
In the general case the components of $I, S, T$ depend smoothly on coordinates
on ${\cal M}$.
The endomorphism 
$I$ endows ${\cal M}$ with the structure of a 
two-dimensional complex manifold,
and $S$ and $T$ 
determine a pair of transverse null foliations.
Let $g$ be a  metric of signature $(2n, 2n)$ 
on $\cal M$.  If $({\cal M}, {\cal J}_{\l})$ is
pseudo-hyper-complex and 
\[
g(TX_1, TX_2)=g(SX_1, SX_2)=-g(IX_1, IX_2)=-g(X_1,X_2)
\]
for all
vectors $X_1, X_2$  then the triple $({\cal M},\; {\cal J}_{\l},\;g)$ 
is called a pseudo-hyper-Hermitian structure.  

From now on we
shall restrict ourselves to oriented four manifolds, where the notions
of  pseudo-hyper-complex and pseudo-hyper-Hermitian structures coincide.
To see it choose any vector $X\in T{\cal M}$, and define a 
conformal structure $[g]$
of signature $(++--)$,
by choosing a conformal frame of vector
fields $(X, IX, SX, TX)$. Any $g\in[g]$ is then pseudo-hyper-Hermitian.
We shall use the following characterisation of the 
pseudo-hyper-Hermiticity condition:
\begin{prop}[\cite{D99}]
\label{Laxmacia}
Let $(X, Y, U, V)$ be four independent real vector fields on 
a four-dimensional real manifold ${\cal M}$, and
let 
\be
\label{hclax}
L_0=X-\lambda V,\;\;\;\;\;
L_1=U-\lambda Y,\;\;\;{\mbox{where}}\;\; 
\lambda \in \CP^1.
\ee
If 
\be
\label{Lemat}
[L_0, L_1]=0
\ee
for every $\lambda$, then $(X, Y, U, V)$ is a null tetrad for a 
pseudo-hyper-Hermitian contravariant metric 
\[
g=X\otimes Y+Y\otimes X-U\otimes V-V\otimes U
\]
on ${\cal M}$. 
Every pseudo-hyper-Hermitian metric
arises in this way.
\end{prop}
For a future reference we write equations (\ref{hclax}) in full:
\be
\label{in_full}
[X, U]=0,\qquad [Y, V]=0,\qquad [X, Y]-[U, V]=0.
\ee 
Given the null tetrad $(X, Y, U, V)$ we 
define the pseudo hyper-complex structure by
\be
\label{hcstructure}
\begin{array}{ccccccccc}
I(X)&=&-V,\qquad I(U)&=&-Y,\qquad I(Y)&=&U,\qquad I(V)&=&X,\\
S(X)&=&V, \qquad S(U)&=&Y,\qquad S(Y)&=&U,\qquad S(V)&=&X,\nonumber\\
T(X)&=&X, \qquad T(U)&=&U,\qquad T(Y)&=&-Y, \qquad T(V)&=&-V.\nonumber
\end{array}
\ee
Proposition \ref{Laxmacia} asserts that
integrability of $I, S, T$ is guaranteed by (\ref{in_full}).
Let $\nu\in\Lambda^4(T^*{\cal M})$ be the volume form on ${\cal M}$.
The covariant metric is conveniently expressed in a dual frame
\begin{eqnarray*}
e_X&=&\nu(..., Y, U, V), \qquad e_Y=\nu(X, ..., U, V), \\
e_U&=&\nu(X, Y, ..., V), \qquad e_V=\nu(X, Y, U, ... ),
\end{eqnarray*}
and is given by
\[
g=e_X\otimes e_Y+e_Y\otimes e_X-e_U\otimes e_V-e_V\otimes e_U.
\]
The result of Boyer \cite{B88} originally formulated for 
hyper-complex manifolds still applies (with some sign alterations) 
in the $(++--)$ signature: a four-manifold is  pseudo-hyper-complex
iff there exists a basis $(\Om_1, \Om_2, \Om_3)$
of the space of self-dual two forms 
$\Lambda^2_+$, and a one-form ${\cal A}$ (called a Lee form)
such that
\be
\label{aaa}
\d\Om_i={\cal A}\wedge \Om_i.
\ee
If we change a representative of a pseudo-conformal structure 
according to $g\rightarrow e^fg$, then
\[
\Om_i\longrightarrow e^f\Om_i,\qquad {\cal A}\rightarrow {\cal A}+
\d f.
\] 
Therefore if ${\cal A}$ is exact, then $g$ is conformally   
pseudo-hyper-K\"ahler (Ricci-flat).  
\section{From the Tzitz\'eica equation to ASDYM}
\label{S3}   
The idea of looking at integrable systems as reductions of the anti-self-dual
Yang-Mills (ASDYM) equations goes back to Ward \cite{Wa85}.
In this section the list of possible reductions will be enlarged by
showing that (\ref{tzitzeica}) arises from the $SL(3, \R)$ ASDYM with
two commuting translational symmetries.
In Subsection \S\S{\ref{s31}} the connection matrices will be reinterpreted
as vector fields on the projective plane.

Consider the flat metric of signature $(2, 2)$ on $\R^4$, which
in double null coordinates $x^a=(x, y, u, v)$ takes a form 
\[
\d x\d y-\d u\d v,
\]
and choose the volume element $\d x\wedge\d y\wedge\d u\wedge\d v$.
Let $A\in T^*\R^4\otimes\mathbf{sl}(3,\R)$ be a connection one-form on a real rank-three vector bundle, and
let $F$ be its curvature two form. In a local trivalisation $A=A_a\d x^a$
and $F=F_{ab}\d x^a\wedge\d x^b$, where
$F_{ab}=[D_a, D_b]$
takes its values in $\mathbf{sl}(3, \R)$.  Here
$D_a=\p_a-A_a$ is the covariant derivative.
The connection is defined up to gauge transformations
$A\rightarrow h^{-1}Ah-h^{-1}\d h$, where $h\in \mbox{Map}(\R^4, SL(3, \R))$.
The ASDYM equations on $A_a$ are $F=-\ast F$, or 
\[
F_{xu}=0, \qquad F_{xy}-F_{uv}=0,\qquad F_{yv}=0.
\]
These equations are equivalent to the commutativity of the Lax pair
\be
\label{LaxSDYM}
L_0=D_x-\l D_v, \qquad L_1=D_u-\l D_y
\ee
for every value of the parameter $\l$.

We shall  require that the connection possess two commuting
translational symmetries $X_1, X_2$ ,
which in our coordinates are in $X_1=\p_u$ and $X_2=\p_v$ directions.
The direct calculation shows that the ASDYM equations are solved by
the following ansatze for Higgs fields $A_u$ and $A_v$, and gauge
fields $A_x$ and $A_y$
\begin{eqnarray}
\label{Tansatz}
A_u&=&-\left (
\begin{array}{ccc}
0&0&0\\
0&0&0\\
e^{\om}&0&0
\end{array}
\right ),\hskip 0.5cm\qquad
A_{v}=-\left (
\begin{array}{ccc}
0&0&1\\
0&0&0\\
0&0&0
\end{array}
\right ),\nonumber\\
A_x&=&-\left (
\begin{array}{ccc}
-\om_x&0&0\\
 1&\om_x&0\\
 0&1&0
\end{array}
\right ),\qquad
A_{y}=-\left (
\begin{array}{ccc}
0&e^{-2\om}&0\\
0&0&e^{\om}\\
0&0&0
\end{array}
\right ),
\end{eqnarray}
iff $\om(x,y)$ satisfies the Tzitz\'eica equation (\ref{tzitzeica}).
We note that 
the reduced Lax pair (\ref{LaxSDYM}) could be obtained directly form 
(\ref{T10}) multiplying the second equation by $\l$.

We connect the ASDYM  equations and those
on a pseudo-hyper-complex four-dimensional metric (\ref{Lemat}) by considering 
gauge potentials that take
values in a Lie algebra of vector fields on some manifold. Proposition
\ref{Laxmacia} reveals one such connection:
We make the identification:
$X=D_{x},\; Y=D_{y},\; U=D_u,\; V=D_{v}$.  By comparing
(\ref{LaxSDYM}) with (\ref{hclax}), we see that the
pseudo-hyper-complex
equation is a reduction of the ASDYM with the infinite-dimensional 
gauge group Diff$({\cal M})$ by
translations along the four coordinate vectors $\p _{x},\; \p_y,
\; \p _{u},\; \p _{v}$.
 
To reveal the connection with the Tzitz\'eica  equation 
we shall proceed in a slightly different way:
Consider the ASDYM equations 
with the gauge
group $G$, being a sup-group of Diff$(\Sm )$, where $\Sm$ is
some two-dimensional real manifold.
We can represent the
components of the connection form of $A$ by  vector fields on $\Sm $ 
depending also on the coordinates on $\R^4$.
Now we suppose that $A$ is invariant under two
translations. The reduced Lax pair will then descend to ${\cal M}=\R^2 \times
\Sm$ and give rise to a pseudo-hyper-complex metric.
A similar idea have been used in \cite{Wa92, DMW98} to construct new classes
of hyper-K\"ahler four-manifolds out of solutions to some integrable 
ODEs and PDEs.

Because we are interested in the case $G=SL(3, \R)$, we take
$\Sm$ to be a real projective plane $\RP^2$ with a natural $PSL(3,
\R)$ group action. The relevant vector fields will be constructed in
the next subsection.
\subsection{$\mathbf{sl}(3, \R)$ as a sub-algebra of $\mathbf{diff}(\RP^2)$}
\label{s31}
To construct a null tetrad for a pseudo-hyper-complex metric 
we will need an explicit embedding $\mathbf{sl}(3, \R)\rightarrow 
\mathbf{diff}(\RP^2)$.
Let 
\[\left (
\begin{array}{ccc}
A_{11}&A_{12}&A_{13}\\
A_{21}&A_{22}&A_{23}\\
A_{31}&A_{32}&A_{33}
\end{array}
\right )\in SL(3, \R).
\]
Consider the projective transformations of a plane with local
coordinates $(p, q)$:
\[
p\longrightarrow\frac{A_{11}p+A_{12}q+A_{13}}{A_{31}p+A_{32}q+A_{33}},\qquad
q\longrightarrow\frac{A_{21}p+A_{22}q+A_{23}}{A_{31}p+A_{32}q+A_{33}}.
\]
This gives rise to a representation of the Lie algebra 
$\mathbf{sl}(3, \R)$ of $SL(3,\R)$
by vector fields on $\RP^2$. The easiest way to obtain this
representation is to consider the infinitesimal linear left  action of $SL(3,\R)$ on
$\R^3$. The generators of this action pushed down to the projective
plane are
\begin{eqnarray*}
&&\p_p,\qquad \p_q,\qquad p\p_q,\qquad q\p_p,\qquad
-p^2\p_p-pq\p_q,\qquad
 -pq\p_p-q^2\p_q,\\
&&p\p_p-q\p_q, \qquad p\p_p+2q\p_q.
\end{eqnarray*}
More precisely, a vector field corresponding to an element
\[
M=\left (
\begin{array}{ccc}
a_{11}&a_{12}&a_{13}\\
a_{21}&-a_{11}-a_{33}&a_{23}\\
a_{31}&a_{32}&a_{33}
\end{array}
\right )\in \mathbf{sl}(3, \R)
\]
is
\begin{eqnarray}
\label{vectorfield}
X_M&=&[a_{13}+  (a_{11}-a_{33})p+a_{12}q-a_{31}p^2-a_{32}pq]\p_p\\
&+&[a_{23}+ a_{21}p-(a_{11}+2a_{33})q-a_{31}pq-a_{32}q^2]\p_q\nonumber.
\end{eqnarray}
\section{Curved metrics from the Tzitz\'eica equation}
\label{S4}
Consider the reduced ASDYM Lax pair (\ref{LaxSDYM})
\[
L_0=\p_x-A_x-\l A_v, \qquad L_1=-A_u-\l(\p_y-A_y),
\]
such that $[L_0, L_1]=0$ yields (\ref{tzitzeica})
and use  (\ref{vectorfield}) to 
 replace the matrices (\ref{Tansatz}) by
vector fields. Now compare the resulting Lax pair
with (\ref{hclax}), and read off the null tetrad for a hyper-complex
metric (some care needs to be taken with signs because 
$[X_M, X_N]=-X_{[M, N]}$). This yields
\begin{eqnarray}
X=\p_x+(-\u_xp+pq)\p_p+(\u_xq-p+q^2)\p_q,&&\qquad
U=-e^\u p^2\p_p-e^\u pq\p_q,\nonumber\\
Y=\p_y-e^{-2\u}q\p_p-e^\u\p_q,{\hskip 3.2cm} &&\qquad
V=\p_p.
\end{eqnarray}
The first two equations in (\ref{in_full}) 
are satisfied trivially, and the third one yields
\[
[X, Y]-[U, V]=(\u_{xy}+e^{-2\u}-e^\u)(p\p_p-q\p_q)
\]
which is $0$ if $\u(x, y)$ satisfies equation (\ref{tzitzeica}).
Let $p=\exp{(P)}, q=\exp{(Q)}$.
The frame of dual one forms is
\begin{eqnarray}
e_X&=&\d x,\qquad e_U=(\u_xe^{-\u-P}+e^{-\u-P+Q}-e^{-\u-Q})\d x-e^{-P-Q}\d
y-e^{-\u-P}\d Q,\\
e_Y&=&\d y,\qquad e_V=(2\u_xe^P-e^{2P-Q})\d x+(e^{Q-2\u}-e^{\u+P-Q})\d y-e^{P}\d Q+e^P\d P.\nonumber
\end{eqnarray}
Finally the metric is given by
\be
\label{Tmetric}
g=2(e_Xe_Y-e_Ue_V).
\ee
It is instructive to verify our calculation by considering the dual formulation
of Boyer.
Using the identification between the two-forms,
and endomorphisms given by $g$ define a basis 
$(\Om_I, \Om_S, \Om_T)$
of $\Lambda^2_+$ by
\[
\Om_I(X_1, X_2)=-g(IX_1, X_2), \qquad
\Om_S(X_1, X_2)=-g(SX_1, X_2), \qquad
\Om_T(X_1, X_2)=-g(TX_1, X_2),
\]
so that
\[
\Om_S=e_X\wedge e_U-e_Y\wedge e_V,
\qquad \Om_T=e_X\wedge e_Y-e_U\wedge e_V, \qquad 
\Om_I=e_X\wedge e_U+e_Y\wedge e_V.
\]
The  Lee form ${\cal A}$ can be found, such that equations
(\ref{aaa}) reduce down to (\ref{tzitzeica}). Indeed, taking
\[
{\cal A}=(3e^{P-Q}-4\u_x)\d x+(3e^{\u-Q}-\u_y)\d y- \d P+2\d Q
\]
yields
\begin{eqnarray*}
&&\d \Om_I-{\cal A}\wedge\Om_I=0,\\
&&\d \Om_S-{\cal A}\wedge\Om_S=0,\\
&&\d \Om_T-{\cal A}\wedge\Om_T=e^\u
[\u_{xy}+e^{-2\u}-e^{\u}]\d x\wedge\d y\wedge\d
(P+Q)=0.        
\end{eqnarray*}
The metric (\ref{Tmetric}) is therefore never conformal to 
pseudo-hyper-K\"ahler because $\d {\cal A}\neq 0$. 
Even the simplest
solution $\u=0$ yields a non-trivial hyper-complex structure\footnote{It is worth remarking that a Tzitz\'eica surface corresponding to $\u=0$ (so called
Jonas Hexenhut) is also 
non-trivial.}
\begin{eqnarray*}
g&=&(e^P-e^{2P-2Q})\d x^2+(3-2e^{P-2Q}-e^{2Q-P})\d x\d y+(e^{-P}-e^{2Q})\d
y^2-2\d Q^2+2\d Q\d P\\
&+&(e^{P-Q}-e^{Q)}\d x\d P+e^{-Q}\d y \d P+(e^Q-2e^{P-Q})\d x\d Q+
(e^{Q-P}-2e^Q)\d y\d Q.
\end{eqnarray*}
The Backlund transformations for the Tzitz\'eica  equation 
\cite{Tz10, BYSS93, CMG99} 
may now be used to generate more complicated metrics.
\section{The twistor correspondence}
\label{S5}
From the point of view of the Yang-Mills equations,  the solutions (\ref{Tmetric}) that we have
obtained are  metrics on the total space of 
${\cal E}$, the  $\RP^2$-bundle  associated to the Yang-Mills bundle.
In this section we explain how our construction ties in with the twistor
correspondences.

Consider the manifold 
 ${\cal Z}=\R^{2, 2}\times \CP^1$ ($\R^{2, 2}$ denotes $\R^4$
with a flat metric of signature $(2, 2)$).
It decomposes into two open sets 
\begin{eqnarray*}
{\cal Z}_+&=&\{ (x^a, \l)\in {\cal Z};\; \mbox{Im}(\l)>0\}={\R}^{2, 2}\times D_+,\\
{\cal Z}_-&=&\{ (x^a, \l)\in {\cal Z}; \;\mbox{Im}(\l)<0\}=\R^{2, 2}\times D_-,
\end{eqnarray*}
where $D_{\pm}$ are two copies of a Poincare disc.
These sub-manifolds are separated by  
\[
{\cal F}_0 =\{ (x^a, \l)\in {\cal Z};\; \mbox{Im}(\l)=0\}={\R^{2, 2}}\times \RP^1.
\]
The complex structures on ${\cal Z}_\pm$ 
are specified by a distribution ${\cal D}$ of anti-holomorphic vector fields
\[
{\cal D}=\{\p_x-\l\p_v,\;\p_u-\l\p_y,\; \p_{\ov{\l}}\}.
\]
The above distribution   with $\l\in\RP^1$
defines a foliation of ${\cal F}_0$
with a quotient ${\cal Z}_0$ which leads to 
a double fibration:
\be
\label{doublefib}
{\cal M}\stackrel{r}\longleftarrow 
{\cal F}_0\stackrel{s}\longrightarrow {\cal Z}_0.
\ee
The {\em twistor space}  ${\cal Z}$ is a three complex dimensional 
union of two open subsets
${\cal Z}_\pm$ 
separated by a three-dimensional real boundary
({\em real twistor space}) ${\cal Z}_0:=s({\cal F}_0)$.

Each point ${\bf x}\in\R^{2, 2}$ determines a holomorphic curve 
 $L_{\bf x}$ made up of two sheets $D_\pm$ of complex structures (\ref{cpx})
compactified by adding  $S^1$:
\[
{\bf x}=(x, y, u, v)\longrightarrow L_{\bf x}=\{(\om^0, \om^1,
 \l):\om^0(\l)=v+\l x,\; \om^1(\l)=u+\l y,\; \l\in\CP^1\}
\]
The normal bundle  $N=T{\cal Z}|_{L_{\bf x}}/T{L_{\bf x}}$
of $L_{\bf x}$
in $\cal Z$ is a direct sum 
of two line bundles with a Chern class equal 
to one ${\cal O}(1)\oplus {\cal O}(1)$. If ${\bf x}$ and ${\bf x'}$ both
lie on a self-dual null plane in $\R^{2, 2}$ then $L_{\bf x}$ and
$L_{{\bf x'}}$ intersect in $\cal Z$ at one point for which $\l \in 
\RP^1$.

Now we turn to the $SL(3, \R)$ ASDYM equations on $\R^{2,2}$
with two commuting symmetries $X_1, X_2$. Let ${\cal E}=\R^{2, 2}\times \RP^2$
be the bundle  associated to the Yang-Mills bundle by the
representation of $SL(3, \R)$ as projective transformations
 of ${\RP^2}$.
 The $SL(3,
\R)$ ASDYM connection defines, by a $(++--)$ version of a 
Ward construction \cite{Wa77}, two
holomorphic vector bundles ${{E_W}_\pm}\rightarrow{\cal Z}_{\pm}$. 
The following construction describes also the general case of
$G=$Diff$({\RP^2})$.  
For this it is convenient to 
use the bundles ${{\cal E}_W}_{\pm}$  associated to ${E_W}_{\pm}$ by the
$G$ action on ${\RP^2}$ (the Ward bundles have
infinite-dimensional fibres).  

On the other hand, any pseudo-hyper-complex four-metric corresponds to a deformed
twistor space ${\cal Z}_{{\cal M}}$, \cite{B88, D99}. 
\begin{prop}
\label{Twmaci}
Let ${\cal Z}_{\cal M}$ be a    
three-dimensional complex manifold with 
\begin{itemize}
\item a four parameter family of
rational curves with normal bundle ${\cal O}(1)\oplus {\cal O}(1)$,
\item a holomorphic projection $\mu:{\cal Z}_{\cal M}\longrightarrow \CP^1$,
\item  an
anti-holomorphic involution 
$\rho:{\cal Z}_{\cal M}\rightarrow{\cal Z}_{\cal M}$ fixing a real equator of  
each rational curve.
\end{itemize}
Then the real moduli space ${\cal M}$ of the $\rho$-invariant 
curves is equipped with 
conformal class $[g]$ of  pseudo-hyper-Hermitian metrics.
Conversely, given a real analytic  
pseudo-hyper-Hermitian metrics
there exists a corresponding twistor space with the above
structures.
\end{prop}
The existence of the holomorphic projection $\mu$ reflects the fact that
the Lax pair (\ref{hclax}) for the pseudo-hyper-complex equations
doesn't contain vector fields $\p_\l$.

In this paper we
have explained how the quotient $q$ of ${\cal E}$ by lifts of $X_1, X_2$ is,
by Proposition \ref{hclax}, equipped with a 
pseudo-hyper-complex
metric .  To give a
more complete picture we can construct the deformed twistor space
directly from ${{\cal E}_W}_{\pm}$ and show that this is the twistor space
of ${\cal M}$.  

Given an analytic
solution to (\ref{tzitzeica}) one can obtain the corresponding
twistor space by equipping ${\cal M}\times \CP^1$ with a structure
of a complex  manifold
${\cal Z}$: The basis of $[0, 1]$ vectors is 
given by the distribution ${\cal D}_{\cal M}$ consisting of 
the Lax pair for the Tzitz\'eica equation 
together with
the standard complex structure 
on the  $\CP^1$.
The point is that this distribution can be obtained directly from
$\cal D$. To see it consider the following chain of correspondences:
\[
\begin{array}{cccccccc}
{{\cal Z}_{\cal M}}={{\cal Z}_{\cal M}}_-\cup
{{\cal Z}_{\cal M}}_0\cup
{{\cal Z}_{\cal M}}_+
&\stackrel{\tilde{\kappa}}\longleftarrow&
{{\cal E}_W}= \R^{2,2}\times\RP^2\times\CP^1
&\stackrel{\pi}\longrightarrow &\{{\cal Z}, {\cal D}\}\nonumber\\
\Big\downarrow & &\Big\uparrow
& &\Big\downarrow\\
{\cal M} &\stackrel{\kappa}\longleftarrow&
{\cal E}=\R^{2,2}\times\RP^2 &\longrightarrow &\R^{2,2}.\nonumber
\end{array}
\]
Here ${\cal Z}$ and ${\cal Z}_{\cal M}$ are the twistor spaces of
$\R^{2, 2}$ and $\cal M$ respectively. 
The twistor space ${{\cal Z}_{\cal M}}$ is defined as 
the quotient $\tilde{\kappa}$ of ${\cal E}_W$ 
by lifts of symmetries $X_1, X_2$.
The complex structures on ${{\cal Z}_{\cal M}}_{\pm}$ are given
a sub-bundle
\[
{\cal D}_M=\tilde{\kappa}(\pi^*{\cal D})=
\{L_0, L_1,  \p_{\ov\l}\}
\subset T{\cal Z}_{\cal M}, 
\]
where  
\begin{eqnarray*}
L_0&=&\p_x+(-\u_xp+pq)\p_p+(\u_xq-p+q^2)\p_q-\l\p_p\\
L_1&=&-e^\u p^2\p_p-e^\u pq\p_q-\l(\p_y-e^{-2\u}q\p_p-e^\u\p_q).
\end{eqnarray*}
Here $\pi$ is a holomorphic fibration of the associated Ward bundle.
The real three-dimensional surface ${{\cal Z}_{\cal M}}_0\subset
{{\cal Z}_{\cal M}}$ is a quotient of 
$\R^{2,2}\times\RP^2\times\RP^1$ by the four-dimensional real distribution
$\{L_0, L_1 , X_1, X_2\}$.
Moreover  ${\cal Z}_{\cal M}$ 
is holomorphicaly  fibered over $\CP^1$ and it 
has a $\O(1)\oplus\O(1)$ rational curve embedded in it.
Both structures are pulled back from ${\cal Z}$  and projected by 
$\tilde{\kappa}$.
The compatibility of these projections 
is a consequence of the
commutativity of the the above diagram, which 
follows  from the integrability the the distribution spanned by 
(lifts of) 
\[
X_1,\; X_2,\; L_0,\; L_1,\; \p_{\ov\l}
\]
 and from the fact that
$(X_1, X_2)$ commute with $(L_0, L_1)$.

\section{Acknowledgements}
I would like to thank Eugene Ferapontov for drawing my attention to
the role of the Tzitz\'eica equation in the theory of surfaces
in affine geometry. I am also grateful to Ian Strachan for having 
read the manuscript carefully.


\begin{thebibliography}{jafsdl}
\frenchspacing 
\bibitem{B88} Boyer, C. A note on hyperhermitian
four-manifolds, 
Proc. Amer. Math. Soc. {\bf 102}, 157-164, (1988).

\bibitem{BYSS93} Boldin, A. Yu. Safin, S. S. Sharipov, R. A. 
On an old article of Tzitzeica and the inverse scattering method. 
J. Math. Phys. {\bf 34} , no. 12, 5801-5809 (1993). 

\bibitem{CMG99} Conte, R., Musette, M.\& Grundland, A. 
Bäcklund transformation of partial differential equations from the Painlevé-Gambier classification. II. Tzitzéica equation. 
J. Math. Phys. {\bf 40} no. 4,  2092--2106 (1999). 

\bibitem{DB77} Dodd, R. K. \& Bullough, R. K. 
 Polynomial conservfed densities for the sine-Gordon equations, 
Proc.~Roy.~Soc.~London A {\bf 352}  481--503 (1977).


\bibitem{D99} Dunajski, M.  The Twisted Photon
Associated to  Hyperhermitian Four Manifolds,  
J. Geom. Phys. {\bf 30}, 266-281 (1999).



\bibitem{DMW98} Dunajski, M. Mason, L.J. \& Woodhouse, N.M.J. 
From 2D Integrable Systems to Self-Dual Gravity, 
J. Phys. A: Math. Gen {\bf 31}, 6019 (1998).

\bibitem{G84} Gaffet, B. 
A class of 1-d gas flows soluble by the inverse scattering transform,
Physica {\bf 26}, 123-131 (1984).



\bibitem{GS99} Grant, J.D.E\ \& Strachan, I.A.B.\ 
Hypercomplex Integrable Systems, 
Nonlinearity {\bf 12}, 1247-1261 (1999).

\bibitem{J53} Jonas, H. Die Differentialgleichung der Affinsphären in einer neuen Gestalt,
Math. Nachr {\bf 10} 331--352 (1953)

\bibitem{M81} Mikhailov, A.V. 
The reduction problem and the inverse scattering method,
Physica D {\bf 3} 73-117 (1981).


\bibitem{Tz08} Tzitz\'eica, G. 
Sur une nouvelle classe de surfaces, 
Rendiconti del Circolo Matematico di Palermo {\bf 25} 180--187 (1908). 


\bibitem{Tz10} Tzitz\'eica, G. 
Sur une nouvelle classe de surfaces,
C. R. Acad. Sci. Paris
{\bf 150} 955--956 (1910).  



\bibitem{Wa77} Ward, R.S.  On self-dual gauge fields, Phys.
Lett. {\bf 61A}, 81-82 (1977).

\bibitem{Wa85} Ward, R.S.  Integrable and solvable systems and
relations among them, Phil. Trans. R. Soc. A {\bf 315}, 451-457 (1985).

\bibitem{Wa92} Ward, R.S.  Infinite-dimensional
gauge groups and special nonlinear gravitons,
J. Geom. Phys., {\bf 8}, 317-325 (1992).


\end{thebibliography}
\end{document}